\documentclass[epj,nopacs]{svjour}
\usepackage{amsmath}
\usepackage{amssymb}
\usepackage{latexsym}
\usepackage[dvips]{graphicx}
\usepackage[english]{babel}

\def\bge{\begin{equation}}
\def\ene{\end{equation}}
\def\bgea{\begin{eqnarray}}
\def\enea{\end{eqnarray}}

\def\bge{\begin{equation}}
\def\ene{\end{equation}}
\def\bgea{\begin{eqnarray}}
\def\enea{\end{eqnarray}}

\def\ls{\raise 1.5pt\hbox{$\,<\;$}\kern -10.5pt\lower3.5pt
          \hbox{$\sim$}\kern 1.5pt} 
\def\gs{\raise 1.5pt\hbox{$\,>\,$}\kern -9.5pt\lower3.5pt
          \hbox{$\sim$}\kern 1.5pt} 

\usepackage{color}

\usepackage{ulem} 

\begin{document}
\sloppy
\title{Quantum Fluctuations at the Planck Scale}
\author{Fulvio Melia\thanks{John Woodruff Simpson Fellow.}}
\institute{Department of Physics, the Applied Math Program, and Department of Astronomy, \\
              The University of Arizona, Tucson, AZ 85721,
              \email{fmelia@email.arizona.edu}}

\authorrunning{Melia}
\titlerunning{Quantum Fluctuations at the Planck Scale}

\date{February 18, 2019}

\abstract{The recently measured cutoff, $k_{\rm min}=4.34\pm0.50/r_{\rm cmb}$ (with
$r_{\rm cmb}$ the comoving distance to the last scattering surface), in the
fluctuation spectrum of the cosmic microwave background, appears to disfavor slow-roll
inflation and the associated transition of modes across the horizon. We show in this
{\it Letter} that $k_{\rm min}$ instead corresponds to the first mode emerging out
of the Planck domain into the semi-classical universe. The required scalar-field
potential is exponential, though not inflationary, and satisfies the zero active mass
condition, $\rho_\phi+3p_\phi=0$. Quite revealingly, the observed amplitude of the
temperature anisotropies requires the quantum fluctuations in $\phi$ to have classicalized
at $\sim 3.5\times 10^{15}$ GeV, consistent with the energy scale in grand unified theories.
Such scalar-field potentials are often associated with Kaluza-Klein cosmologies, string
theory and even supergravity.}
\maketitle

\section{Introduction}
Three major satellite missions have now confirmed the lack of angular correlation
at $\theta\gtrsim 60^\circ$ in the cosmic microwave background (CMB)
\cite{Wright:1996,Bennett:2003,Planck:2018}, contrasting with the predictions of
standard inflationary cosmology \cite{Guth:1981,Linde:1982,Mukhanov:1992,Copi:2009}.
Though many possible instrumental and observational selection effects have been
considered as the cause of this deficiency, today we appear to be left solely with
cosmic variance as an explanation for the missing correlations. But even this
conclusion disfavors the conventional picture at $\gtrsim 3\sigma$. A recent
in-depth analysis of the {\it Planck} measurements \cite{MeliaLopez:2018}
reveals that a far more likely reason for the missing angular correlations is a
non-zero minimum wavenumber, $k_{\rm min}$, in the fluctuation power spectrum $P(k)$.
This evidence now shows quite compellingly that a zero $k_{\rm min}$ is ruled out at a
level of confidence {\it exceeding} $8\sigma$. The {\it Planck} data suggest, instead, a
cutoff $k_{\rm min}=4.34\pm0.50/r_{\rm cmb}$, where $r_{\rm cmb}$ is the comoving distance
to the surface of last scattering.

This measurement disfavors basic slow-roll inflation because, in the standard picture,
$k_{\rm min}$ would have been the first mode crossing the horizon during the
near-exponential expansion \cite{LiuMelia2019}. Thus, for an assumed inflaton potential
$V(\phi)$, $k_{\rm min}$ would have signaled the precise time at which inflation started.
As it turns out, however, one could not simultaneously solve the temperature horizon problem
and account for the observed fluctuation spectrum in the CMB with this minimum cutoff. If
inflation is to work, the sequence of steps preceding (or succeeding) the inflationary
phase would necessarily have to be more complicated than has been conjectured thus far.

In this {\it Letter}, we take an alternative approach, and simplify the concept of
how a scalar field ought to have behaved in order to comply with the observational
constraints. Arguably, the most significant, relevant measurements we have to date
are (1) the scalar spectral index, $n_s=0.9649\pm 0.0042$, in the power spectrum
$P(k)=A_s(k/k_0)^{n_s-1}$ \cite{Planck:2018}, (2) the amplitude $A_s=(2.1\pm 0.04)
\times 10^{-9}$ \cite{Planck:2018}, and now (3) the hard cutoff $k_{\rm min}=
4.34\pm0.50/r_{\rm cmb}$ \cite{MeliaLopez:2018}. To be clear, none of these measurements
argues {\it against} the influence of a scalar field, or anisotropies arising from its quantum
fluctuations (QFs), but there are now good reasons to question whether its potential
is truly inflationary.

Indeed, a significant body of evidence is accumulating in favor
of a zero active mass ($\rho+3p=0$) equation-of-state in the cosmic fluid, based
on the analysis of over 25 different kinds of observation at low and high
redshifts. A summary of these results may be found in Table~2 of
ref.~\cite{Melia:2018a}. A notable feature of such a universe is that it lacks
a horizon problem \cite{Melia:2013,Melia:2018b}, so our failure to build a fully
successful inflationary paradigm may simply reflect the fact that the universe
does not need it. This is the key assumption we shall make in this {\it Letter}.
Some additional support for such a cosmological model, based on
an alternative theoretical concept, may also be found in ref.~\cite{John:2000}, 
with an updated discussion in ref.~\cite{John:2019}.

In our step-by-step comparison of the roles played by a non-inflaton $\phi$ versus
a conventional inflaton field, we shall distinguish the former from the latter by
informally calling it a `numen' field, representing the earliest manifestation of
substance in the universe, with an equation-of-state $\rho_\phi+3p_\phi=0$. We shall
readily understand why the numen field appears to be superior to an inflaton field in
accounting for the measured values of $n_s$, $A_s$ and $k_{\rm min}$. To do
so, let us first derive the perturbation growth equation for zero active mass, and
see how a numen QF evolves as the universe expands. The essential steps are by now
well known, having appeared in both the primary and secondary literature, so we here
show only the key results and refer the reader to several other influential publications
for the details.

The perturbed Friedmann-Lema\^itre-Robertson-Walker (FLRW) spacetime for the
linearized scalar fluctuations is given by the line element
\cite{Bardeen:1980,Kodama:1984,Mukhanov:1992,Bassett:2006}
\begin{eqnarray}
ds^2 &=& (1+2A)\,dt^2-2a(t)(\partial_iB)\,dt\,dx^i-\nonumber\\ 
&\null& a^2(t)\left[(1-2\psi)\delta_{ij}+2(\partial_i
\partial_jE)+h_{ij}\right]\,dx^i\,dx^j\,,\quad
\end{eqnarray}
where indices $i$ and $j$ denote spatial coordinates, $a(t)$ is the expansion
factor, and $A$, $B$, $\psi$ and $E$ describe the scalar metric perturbations,
while $h_{ij}$ are the tensor perturbations.

For small perturbations about the homogeneous numen field $\phi_0(t)$,
\begin{equation}
\phi(t,\vec{x}) = \phi_0(t)+\delta\phi(t,\vec{x})\;,
\end{equation}
one can identify in the comoving frame the curvature perturbation $\Theta$ on
hypersurfaces orthogonal to comoving worldlines \cite{Bardeen:1980} as a gauge
invariant combination of the metric perturbation $\psi$ and the scalar field
perturbation $\delta\phi$:
\begin{equation}
\Theta\equiv \psi+\left({H\over \dot{\phi}}\right)\,\delta\phi\;.
\end{equation}
Then, expanding $\Theta$ in Fourier modes, and inserting the linearized metric
(Eq.~1) into Einstein's equations, one derives the perturbed equation of motion
\begin{equation}
\Theta_k^{\prime\prime}+2\left({z^\prime\over z}\right)\Theta_k^\prime+k^2\Theta_k=0\;,
\end{equation}
where overprime denotes a derivative with respect to conformal time
$d\tau=dt/a(t)$, and
\begin{equation}
z\equiv {a(t)(\rho_\phi+p_\phi)^{1/2}\over H}\;.
\end{equation}

Since $\phi_0$ is homogeneous, the energy density $\rho_\phi$ and
pressure $p_\phi$ are simply given as
\begin{equation}
\rho_\phi={1\over 2}{\dot{\phi}}^2+V(\phi)\;,
\end{equation}
and
\begin{equation}
p_\phi={1\over 2}{\dot{\phi}}^2-V(\phi)\;.
\end{equation}
The zero active mass condition $\rho_\phi+3p_\phi=0$ therefore constrains the
potential to have the form $V(\phi)={{\dot{\phi}}^2}$, with the explicit solution
\begin{equation}
V(\phi)=V_0\,\exp\left\{-{2\sqrt{4\pi}\over m_{\rm P}}\,\phi\right\}\;.
\end{equation}
The numen field is therefore a special member of the category of minimally coupled
fields explored in the 1980's, designed to produce so-called power-law inflation
\cite{Abbott:1984,Lucchin:1985,Barrow:1987,Liddle:1989}. Unlike the other fields,
however, the numen field's zero active mass equation-of-state renders it the sole
member of this class that actually does {\it not} inflate, since $a(t)=t/t_0$, in
terms of the age of the universe, $t_0$. (This normalization of the expansion
factor is possible as long as the FLRW metric is spatially flat, which all the
observations strongly suggest.)

It is straightforward to see that the conformal time may therefore be written
$\tau(t)=t_0\ln a(t)$, when the zero of $\tau$ is chosen to coincide with $t=t_0$.
Thus, $z=m_{\rm P}a(t)/\sqrt{4\pi}$, in terms of the Planck mass
$m_{\rm P}\equiv 1/\sqrt{G}$, so that $z^\prime/z=1/t_0$ and $z^{\prime\prime}/z=1/t_0^2$.
If we now follow the conventional approach of rewriting Equation~(4) using the
Mukhanov-Sasaki variable $u_k\equiv z\Theta_k$, the resulting curvature perturbation
equation becomes
\begin{equation}
u_k^{\prime\prime}+\alpha_k^2 u_k=0\;,
\end{equation}
where
\begin{equation}
\alpha_k\equiv {1\over t_0}\sqrt{\left(2\pi R_{\rm h}\over \lambda_k\right)^2-1}\;,
\end{equation}
and $\lambda_k\equiv 2\pi a(t)/k$ is the proper wavelength of mode $k$. The quantity
$R_{\rm h}\equiv c/H=ct$ is the apparent (or gravitational) radius \cite{Melia:2018c}
which, as one can see, coincides with the Hubble horizon in a spatially flat universe.
The frequency $\alpha_k$ is critical to understanding how and why QFs
in the numen field are far better suited to the above measurements than those in an
inflaton field.

This advantage may be realized first and foremost by considering the explanation provided
by Equation~(10) for the origin of $k_{\rm min}$. The most significant departure
of this frequency from its inflaton counterpart is that both $R_{\rm h}$
and $\lambda_k$ scale exactly the same way with time. The ratio $R_{\rm h}/\lambda_k$
or, equivalently, $kR_{\rm h}/a(t)$, is therefore constant for each mode $k$. Numen QFs
do not cross back and forth across the horizon, so once the
wavelength of a mode is established upon exiting into the semi-classical universe,
it remains a fixed fraction of $R_{\rm h}$ as they expand with time.

The growth Equation~(9) has analytic solutions
\begin{equation}
u_k(\tau) = \left\{ \begin{array}{ll}
         B(k)\,e^{\pm i\alpha_k\tau} & \mbox{($2\pi R_{\rm h}>\lambda_k$)} \\
         B(k)\,e^{\pm |\alpha_k|\tau} & \mbox{($2\pi R_{\rm h}<\lambda_k$)}\end{array} \right. \;,
\end{equation}
showing that all modes $u_k$ with a wavelength smaller than $2\pi R_{\rm h}$ oscillate,
while the super-horizon ones do not, echoing some of the physical characteristics we encounter
with a traditional inflaton field. But here the mode with the longest wavelength relevant
to the formation of structure is therefore always the one for which $\lambda_k(t)=2\pi R_{\rm h}(t)$,
which happens to be $k_{\rm min}^{\rm N}=1/t_0$. (In spite of its appearance, this equality
does {\it not} constitute a coincidence, because $a(t)$ is itself normalized to $1$ at $t_0$.)

We shall now show that $k_{\rm min}$ clearly represents the emergence of the first QF mode
out of the Planck regime by interpreting it as a measure of $k_{\rm min}^{\rm N}$.
Defining the Planck scale as the length $\lambda_{\rm P}$ for which the Compton wavelength
$\lambda_{\rm C}\equiv 2\pi/m$ for mass $m$ equals its Schwarzschild radius
$R_{\rm h}\equiv 2Gm$, we find that $\lambda_{\rm P}\equiv \sqrt{4\pi G}$. Given
that $\lambda_{\rm C}$ grows as $R_{\rm h}$ shrinks, quantum mechanics does not
actually offer any clear insight into how one should handle modes with wavelengths
shorter than $2\pi\lambda_{\rm P}$ (the factor $2\pi$ arising from the definition
of $\lambda_{\rm P}$ in terms of $R_{\rm  h}$). This is a serious problem for inflationary cosmology
because the fluctuation amplitude $A_s$ observed in the CMB requires QF modes to have
been seeded well before the Planck time $t_{\rm P}=\lambda_{\rm P}$, giving rise to
what is commonly referred to as the ``Trans-Planckian Problem" \cite{Martin:2001}.

But a numen field can avoid this fundamental inconsistency altogether if we argue
that each mode $k$ emerges into the semi-classical universe when its wavelength
$\lambda_k$ equals $2\pi\lambda_{\rm P}$, after which it evolves according to the
oscillatory solution in Equation~(11). As we shall see, the fact that each succeeding
$k$ emerges at later times then naturally produces the near scale-free power spectrum $P(k)$
with $n_s\sim 1$. The notion that modes may have been born at a particular spatial
scale has already been considered in other guises by several other authors, notably
Hollands and Wald \cite{Hollands:2002} though, in their case, the fundamental scale
was not related to $\lambda_{\rm P}$. Other considerations supporting this concept
had also been presented in refs.~\cite{Brandenberger:2002,Hassan:2003}.

For the numen field, however, we know that the fundamental scale has to be $\lambda_{\rm P}$
for the following simple---though quite compelling---reason. From the expression
$k=2\pi a(t)/\lambda_k(t)$, we see that the observed value of $k_{\rm min}$ defines
the time $t_{\rm min}$ at which the very first mode emerged into the semi-classical
universe. Thus,
\begin{equation}
t_{\rm min}={4.34\,t_{\rm P}\over \ln(1+z_{\rm cmb})}\;,
\end{equation}
where $z_{\rm cmb}$ is the redshift at the last scattering surface. In $\Lambda$CDM,
$z_{\rm cmb}\sim 1000$, for which $t_{\rm min}\sim 0.63 t_{\rm P}$. Of course, we
don't know what the exact value of $z_{\rm cmb}$ will be if we modify the expansion
history, but the value of $t_{\rm min}$ is only weakly dependent on this redshift anyway.
For example, even if we were to adopt an extremely different value $z_{\rm cmb}=50$,
the implied first emergence would have occurred at $t_{\rm min}\sim 1.1 t_{\rm P}$.

Thus, the (spatially) largest numen mode measured in the CMB had to emerge
from the Planck scale {\it at approximately the Planck time}. In other words, the
measured value of $k_{\rm min}$ corresponds to the very first mode that could have
physically exited the Planck domain shortly after the Big Bang. This situation is quite
unique among cosmological scalar fields, inflaton or otherwise, for here the macroscopic
anisotropies in the CMB are shown to be directly coupled to QFs at the Planck
scale, an interpretation of $k_{\rm min}$ that supports the viability of a numen
field as an explanation for the origin of structure in the early universe.

As we explore this concept further, we also uncover a natural mechanism for
generating a near scale-free power spectrum $P(k)$, that does not rely on the
poorly motivated use of a Bunch-Davies vacuum below the Planck scale for the QF
normalization $B(k)$ in Equation~(11) \cite{Bunch:1978}, where wavelengths
shorter than $\lambda_{\rm C}$ are difficult to interpret quantum mechanically.
Nor does this mechanism rely on the uncertain transition of modes back and forth
across the horizon, as needs to happen for an inflaton field.

To calculate the QF power spectrum $P(k)$, it is necessary to establish the
normalization $B(k)$ of the modes, which is typically done by minimizing the
expectation value of the Hamiltonian. But in time-dependent spacetimes, such as
inflationary $\Lambda$CDM, the effects of curvature make the frequencies time
dependent. The solution in that case is to seed the fluctuations in the remote past,
where the observable modes today presumably had a wavelength much smaller than the
horizon. They were therefore unaffected by gravity and could thus be normalized as in
Minkowski space, with $B(k)=1/\sqrt{2k}$ \cite{Bassett:2006}. This approach defines
the Bunch-Davies vacuum \cite{Bunch:1978} though, as noted earlier, it is subject
to a possible trans-Planckian inconsistency.

But this problem completely disappears for the numen field, even though the modes
would have emerged at the Planck scale, for the simple reason that the zero active
mass condition makes the frame into which the modes emerged from the Planck domain be
geodesic. That is, although the Hubble frame was expanding, it was nonetheless in
free fall, with zero internal acceleration. This is confirmed quantitatively by the
fact that the frequencies $\alpha_k$ in Equation~(10) are time-independent, since both
$R_{\rm h}$ and $\lambda_k$ change with time in the same way. We shall therefore
set $B(k)=1/\sqrt{2\alpha_k}$ for the numen QFs.

After time $t_{\rm min}$, which we now understand was effectively $t_{\rm P}$,
modes continued to exit from the Planck domain, but always with a wavelength
$\lambda_k=2\pi\lambda_{\rm P}$. That is, $k=a(t)/\lambda_{\rm P}$, suggesting
that mode $k$ exited at time
\begin{equation}
t_k\equiv t_0\lambda_{\rm P}k\;.
\end{equation}
Then, from Equation~(11) and the definition of $u_k$, we find that
\begin{equation}
|\Theta_k|^2={2\pi\over m_{\rm P}^2}{1\over \alpha_k a^2}\;.
\end{equation}
The power spectrum $P_\Theta(k)$ is defined according to
\begin{equation}
P_\Theta(k)\equiv {k^3\over 2\pi^2}|\Theta_k|^2\;,
\end{equation}
and we therefore see from Equations~(10), (14) and (15) that
\begin{equation}
P_\Theta(k)={1\over (2\pi)^2}\left[{a(t_k)\over a(t)}\right]^2\left\{1-
\left({k_{\rm min}^{\rm N}\over k}\right)^2\right\}^{-1/2}\;.
\end{equation}

What is still missing from this picture is a firm understanding of how
classicalization converts homogeneous, isotropic quantum fluctuations
into classical anisotropies. This is a common problem with all models
invoking a quantum origin for the perturbations, and has not yet been
solved \cite{Penrose:2004,Perez:2006,Mukhanov:2005,Weinberg:2008,Lyth:2009,Bengochea:2015}.
It is very likely, however, that the dynamics of classicalization is associated
with a particular length (or energy) scale, $L_*$, analogous to the Planck scale
$\lambda_{\rm P}$ \cite{Brouzakis:2012,Dvali:2012}, which will here serve the
principal purpose of establishing the amplitude of the quantum fluctuations.

Mode $k$ will reach the classicalization scale at $a(t_k^*)=L_*k/2\pi$,
corresponding to time $t_k^*=t_0L_*k$. And so Equation~(16) may be re-written
\begin{equation}
P_\Theta(k)={1\over (2\pi)^2}\left[{\lambda_{\rm P}\over L_*}\right]^2\left\{1-
\left({k_{\rm min}^{\rm N}\over k}\right)^2\right\}^{-1/2}\;,
\end{equation}
which should then be compared directly with the observed power spectrum
$P(k)=A_s(k/k_0)^{n_s-1}$ in the CMB \cite{Planck:2018}.
We find from the measured value of $A_s$ that $L_*$$\sim$$3.5\times 10^3 
\lambda_{\rm P}$. Therefore, with the Planck scale set at $m_{\rm P}\approx 
1.22\times 10^{19}$ GeV, we infer that classicalization of the numen QFs
must have occurred at roughly $3.5\times 10^{15}$ GeV, remarkably consistent
with the energy scale in grand unified theories (GUTs). Unlike our more
robust conclusion regarding the origin of $k_{\rm min}$, this inference
may be more speculative, given that the physics describing this process
clearly lies beyond the standard model, but it is reasonable to suppose
that the numen QFs oscillated until $t\sim 3.5\times 10^3t_{\rm P}$, at
which point the numen field devolved into GUT particles and the perturbation
amplitude remained frozen thereafter.

From Equation~(17), we also see that the numen scalar curvature perturbations
have an almost scale-free spectrum, as one can see from the standard definition
\begin{equation}
n_s=1+{d\,\ln P_\Theta (k)\over d\,\ln k}=
1-{2\over 2(k/k_{\rm min}^{\rm N})^2-1}\;,
\end{equation}
showing that $n_s$ is slightly less than $1$, and suggesting that the
deviation from a pure scale-free spectrum is due to the difference between
$k$ and $\alpha_k$ in Equation~(10), which ultimately arises from the Hubble
expansion term $\Theta^\prime_k$ in the growth Equation~(4).

The identification of $k_{\rm min}$ with the first QF to have emerged
into the semi-classical universe from the Planck domain shortly after
the Big Bang is very exciting. At this level of sophistication, we do
not yet encounter a trans-Planckian inconsistency though, clearly,
probing the physics prior to $t_{\rm P}$ should be a principal focus
of attempts at understanding the nature of $\phi$ and its emergence
into the semi-classical universe.

Already, we can say that one of the strongest arguments in favor of the
super-horizon freeze-out mechanism during inflation also supports the
numen field emergence out of the Planck domain. The coherence of the
observed CMB fluctuations \cite{Dodelson:2003} requires all of the
Fourier modes of a given wavenumber to have an identical phase. With
an inflaton field, this happens because they all cross the horizon
at the same time. Similarly, all of the numen QFs with a given $k$
have the same phase because they emerge from the Planck domain at
a fixed time $t_k$. The numen mechanism is a little simpler because
it requires fewer steps and, at the same time, appears to avoid at
least some of the difficulties faced by conventional inflation.

Our final conclusion from this discussion is that if the early universe
was indeed dominated by a single numen field, then we know its potential
(Eq.~8) quite precisely. Exponential forms such as these are well
motivated in Kaluza-Klein cosmologies, string theories and even
supergravity. The measurement of $k_{\rm min}$ in the CMB fluctuations
may therefore catalyze further meaningful development in these areas.

{\acknowledgement
I am grateful to Amherst College for its support through a John Woodruff Simpson Lectureship.
\endacknowledgement}

%
%

\end{document}